\documentclass{article}

\usepackage[utf8]{inputenc} % allow utf-8 input
\usepackage[T1]{fontenc}    % use 8-bit T1 fonts
\usepackage{hyperref}       % hyperlinks
\usepackage{url}            % simple URL typesetting
\usepackage{booktabs}       % professional-quality tables
\usepackage{amsfonts}       % blackboard math symbols
\usepackage{nicefrac}       % compact symbols for 1/2, etc.
\usepackage{microtype}      % microtypography
\usepackage{here}           % Okamoto add
\usepackage{graphicx}
\title{Modular decomposition of Markov chain: detecting hierarchical organization of pervasive communities}

\author{
  Hiroshi Okamoto\thanks{HO partly conducted this work at DWANGO Co., Ltd.} \\
  Department of Bioengineering\\
  The University of Tokyo\\
  Tokyo 113-8656 JAPAN\\
  \texttt{okamoto@coi.t.u-tokyo.ac.jp} \\
  \and
 Xu-le Qiu\\
  Research \& Development Group\\
  Fuji Xerox Co., Ltd.\\
  Kanagawa 220-8668 JAPAN\\
  \texttt{qiu.xu-le@fujixerox.co.jp} \\
}
\date{\empty}

\begin{document}
\maketitle

%\begin{abstract}
\noindent
{\bf In network science, a group of nodes connected with each other at higher probability than with those outside the group is referred to as a community. From the perspective that individual communities are associated with functional modules constituting complex systems described by networks, discovering communities is primarily important for understanding overall functions of these systems. Much effort has been devoted to developing methods to detect communities in networks since the early days of network science. Nevertheless, the method to reveal key characteristics of communities in real-world network remains to be established. Here we formulate decomposition of a random walk spreading over the entire network into local modules as proxy for communities. This formulation will reveal the pervasive structure of communities and their hierarchical organization, which are the hallmarks of real-world networks but are out of reach of most existing methods.}
%\end{abstract}

% keywords can be removed

\begin{figure}[ht]
\begin{center}
%\fbox{\rule{60mm}{0mm}\rule{0mm}{40mm}}
\includegraphics[width=115mm]{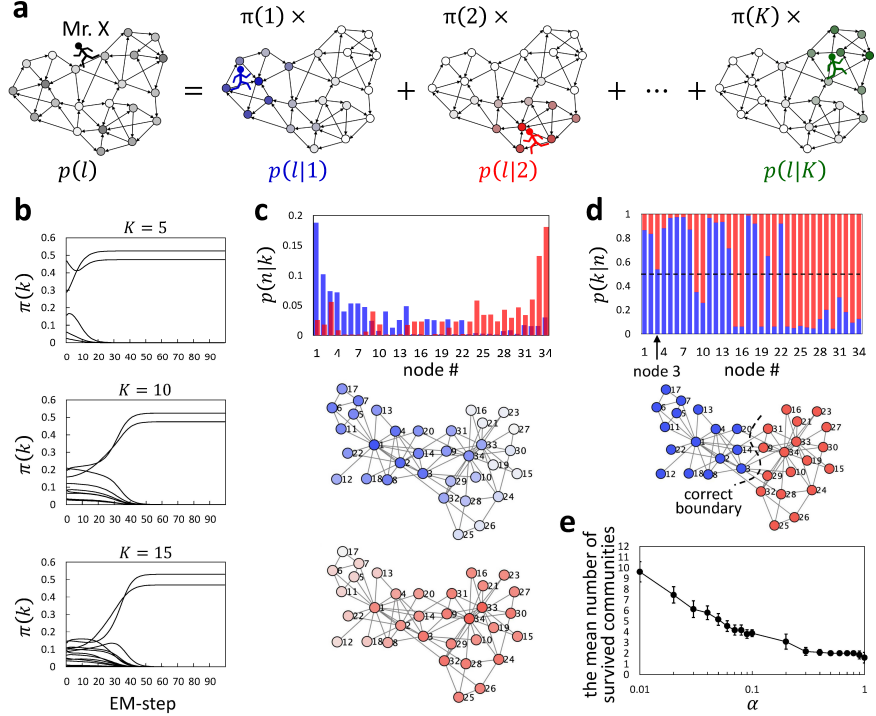}
\caption{
{\small
(a) Schematic drawing of modular decomposition of Markov chain (MDMC). (b) Evolution of community sizes $\pi(k)$ with the EM-setp for different initial setting of the number of communities ($K=5, 10$ or $15$). (c) {\it Top,} the probability distribution $p(n|k)$ delineating the pervasive structure of either community $k$ (indicated by blue or red). {\it Middle and bottom,} network visualization of the pervasive structure of either community. (d) {\it Top,} the relative belonging $p(k|n)$ of each node $n$ to either community $k$. {\it Bottom,} network visualization of the main belonging of each node. (e) The number of survived communities obtained at each value for $\alpha$, averaged over 24 trials, is plotted as a function $\alpha$, the only parameter of MDMC.
}
}
\label{Fig1}
\end{center}
\end{figure}

The idea of exploiting random walk is useful for designing effective and efficient methods for community detection
%%%%% citation
\cite{Lambiotte2009, Delvenne2010, Mucha2010, Rosvall2007, Rosvall2008, Shaub2012, Kheirkhahzadeh2016}. 
%%%%%
Our formulation, which we call modular decomposition of Markov chain (MDMC), is in line with this idea. 
The intuition of MDMC is illustrated in Fig.\ref{Fig1}a. Suppose a random walker, say Mr. X, travelling in the network. The probability that Mr. X is moving along link $l$ is given by 
\begin{equation}
p(l) = T_{n_l m_l} p(m_l)\ ,
\label{Eq_link_prob}
\end{equation}
where $m_l$ and $n_l$ denote the initial- and terminal-end nodes of link $l$, respectively; $T_{nm}$ is the rate for transition from node $m$ to node $n$; $p(n)$ is the probability that Mr. X is at node $n$, which is given in the stationary state of the Markov chain
\begin{equation}
p_t(n) = \sum_{m=1}^{N}T_{nm} p_{t-1}(m)\ ,
\label{Eq_Markov_chain}
\end{equation}
where $N$ is the total number of nodes. 

Given that the network comprises several communities, Mr. X is trapped by some community and stays there for a while; at some time, he chances to move to another community and then stays there for a while, and so forth. The probability that Mr. X is moving along link $l$ conditioned that he is staying in community $k$ is modelled by the products of categorical distributions: 
\begin{equation}
p(l|k) = \prod_{n=1}^{N}\left[ p(n|k) \right]^{\delta_{n,\ n_l}}
\times \prod_{m=1}^{N}\left[ p(m|k) \right]^{\delta_{m,\ m_l}}
=p(n_l|k)p(m_l|k)\ ,
\label{Eq_link_prob_cond_by_community}
\end{equation}
where $p(n|k)$ is the probability that Mr. X is at node $n$ conditioned that he is staying in community $k$. 

The `global' probability $p(l)$ will then be expressed as a mixture of `local' probabilities $p(l|k)$:  
\begin{equation}
p(l) = \sum_{k=1}^{K} \pi(k) p(l|k)\ ,
\label{Eq_MDMC}
\end{equation}
where $K$ is the putative number of communities; $\pi(k)\ (\ge 0)$ is the probability that Mr. X is staying in community $k$ and satisfies $\sum_{k=1}^{K} \pi(k) =1$. Community detection is attained if Eq (\ref{Eq_MDMC}) is solved to $p(l|k)$ (or $p(n|k)$) and $\pi(k)$. Indeed, one can derive the expectation-maximization (EM) algorithm to solve these (see Methods for detailed derivation), which is given in the form:

\noindent
{\it E-step}
\begin{equation}
r(k|l) = \frac{\pi_{t-1}(k) p_{t-1}(n_l|k) p_{t-1}(m_l|k)}
{\sum_{k=1}^{K} \pi_{t-1}(k) p_{t-1}(n_l|k) p_{t-1}(m_l|k)}\ ,
\label{Eq_E_step}
\end{equation}
{\it M-step}
\begin{equation}
\pi_{t} (k)= \sum_{l=1}^{L} p(l)r(k|l)\ ,
\label{Eq_M_step_pi}
\end{equation}
\begin{eqnarray}
p_{t}(n|k) &=& \frac{\alpha}{\alpha + \pi_{t}(k)} \sum_{m=1}^{N} T_{nm}p_{t-1}(m|k) 
\nonumber \\
&+& \frac{1}{\alpha + \pi_{t}(k)} \frac{1}{2}
\sum_{l=1}^{L} p(l)r(k|l) \left( \delta_{n,\ n_l} + \delta_{n,\ m_l} \right)\ ,
\label{Eq_M_step_p}
\end{eqnarray}
where $L$ is is the total number of links; $\alpha$ is the only parameter of MDMC, which will turn out to be controlling the resolution of community detection. 

Node $n$'s rating in community $k$ is defined by $p(n|k)$, which takes a continuous value ranging form 0 to 1. This means that any node is a member of any community irrespective of its rating in that community. In other words, there are no clear boundaries that separate nodes (or links) inside and outside communities. Such structure of communities is hence described as ``pervasive'' 
%%%%% citation
\cite{Fortunato2016}. 
%%%%%
The probability that Mr. X is staying in community $k$ conditioned that he is at node $n$ is given by the Bayes formula
\begin{equation}
p(k|n) = \frac{p(n|k)\pi(k)}{\pi(k)}\ ,
\label{Eq_Bayes}
\end{equation}
which represents the relative ``belonging'' of node $n$ to community $k$. To compare communities detected by MDMC with ground-truth communities of benchmark networks, which are commonly presented as sets of non-overlapping communities called ``partitions'' 
%%%%% citation
\cite{Fortunato2016}, 
%%%%%
we define the community to which node $n$ mainly belongs by 
$\mathrm{arg\ max}_k p(k|n)$. 

MDMC thus detects communities as pervasively structured objects. Indeed, the pervasive structure is a key feature of communities in a variety of real-world networks. For instance, the cell-assembly hypothesis
%%%%% citation
\cite{Hebb1949}, 
%%%%%
a leading principle of modern neuroscience, states that neurons that are frequently co-activated tend to connect each other, thereby forming densely connected structures called ``cell assemblies'', or in the words of network science, ``communities''. Cell assemblies are associated with functional modules of parallel distributed processing of the brain, implying that they are pervasively structured. Also, communities in social networks would be originally pervasive; each person would belong to various communities with variable degree of participation. Artificial boundaries that separate members and non-members would be set only after the formation of a faction is declared or a list of names is established. 

First, we demonstrate fundamental properties of MDMC using Zachary’s karate club network 
%%%%% citation
\cite{Zachary1977, Girvan2002}. 
%%%%%
Figure \ref{Fig1}b shows how the probabilities $\pi(k)$ of communities evolve with the EM step. For any $K$ chosen initially, $\pi(k)$ of only two communities survive settling at positive values, whereas those of the $K-2$ others decay to zero (for the current choice of the parameter value, here $\alpha=0.5$). Thus, MDMC automatically determines the final number of communities. Figure \ref{Fig1}c shows the probability distributions $p(n|k)$ for the survived communities, which delineate their pervasive structures. For most nodes the relative belonging $p(k|n)$ to either community is near (but not exactly) unity or zero, but for node 3 they are very close (Fig. \ref{Fig1}d, top). Indecisive belonging of node 3 to either community, discovered by our pervasive community detection, accounts for why this node is often misclassified in conventional community detection
%%%%% citation
\cite{Peel2017}. 
%%%%%
Identifying the main belonging of each node correctly recovers the actual separation of the karate club, which is the commonly used ground truth of this best-known social network (Fig. \ref{Fig1}d, bottom). We also addressed the role for $\alpha$, the only parameter of MDMC. The number of survived communities decreased as  $\alpha$ increased (Fig. \ref{Fig1}e). Thus, parameter $\alpha$ controls the resolution of decomposition of the network into communities: For smaller $\alpha$, the network is decomposed into more communities of smaller sizes. 
%
%%%%%%%%%%%%%%%%%%%% Figure 1
%%%%%%%%%%%%%%%%%%%%
%

\begin{figure}[H]
\begin{center}
%\fbox{\rule{60mm}{0mm}\rule{0mm}{40mm}}
\includegraphics[width=115mm]{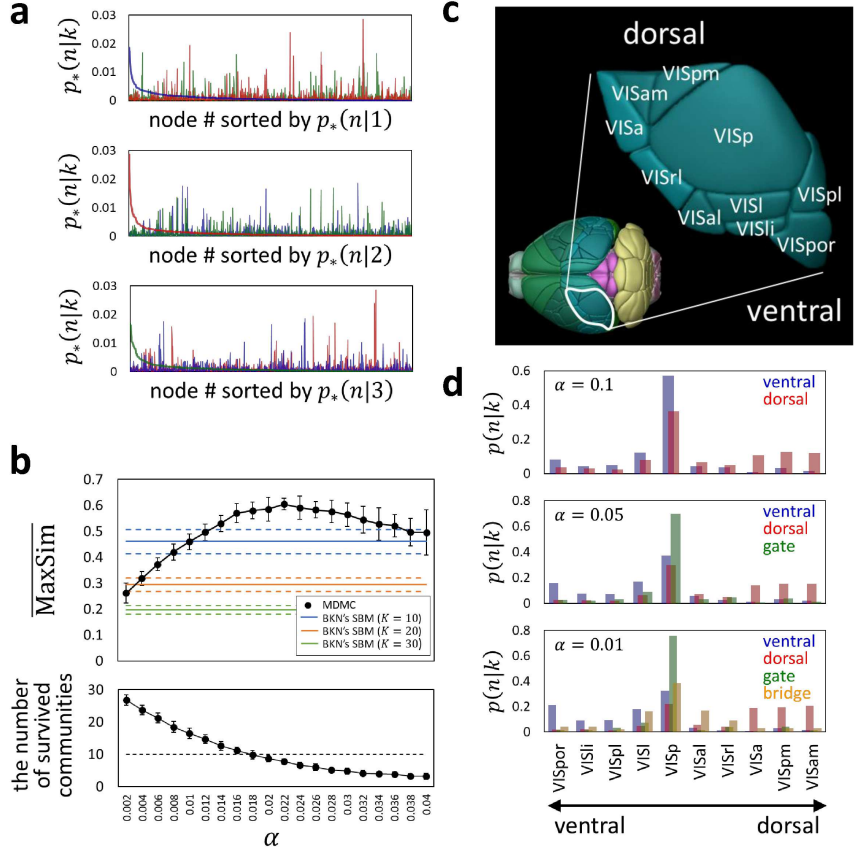}
\caption{
{\small 
(a) The probability distributions $p_*(n|k_*)$ for three planted communities ($k_*=1, 2$ and $3$), out of $K_*=10$ communities, are shown in the top, middle and bottom panels with the node number (\#) sorted according to the descending order of $p_*(n|1)$ (blue), $p_*(n|2)$ (red) and $p_*(n|3)$ (green), respectively. (b) {\it Top,} MaxSim for MDMC averaged over 24 benchmark networks is plotted as a function of $\alpha$ (filled circle). MaxSim by BKN's SBM for $K=$10, 20 or 30 is indicated by each horizontal line. {\it Bottom,} the corresponding number of survived communities plotted as a function of $\alpha$. (c) The mouse visual cortex traced by the Brain Explore (https://mouse.brain-map.org/static/brainexplorer). (d) The probability distributions $p(n|k)$ for communities detected from the network of the 10 areas of the mouse visual cortex for $\alpha=0.1$, $0.05$ and $0.01$ are shown in the top, middle and bottom panels, respectively. 
}
}
\label{Fig2}
\end{center}
\end{figure}

Second, we evaluated MDMC's performance of pervasive community detection. For this, benchmark networks planted with $K_*=10$ pervasive communities were mathematically synthesized as described in Methods. Each planted community $k_*$ is defined by the probability distribution  $p_*(n|k_*)$; Panels in Fig. \ref{Fig2}a show the probability distributions for planted communities $k_*=1, 2$ or $3$ of the same network, but with the node number (\#) sorted in descending order of $p_*(n|1)$, $p_*(n|2)$ or $p_*(n|3)$. The performance was measured by the maximum similarity (MaxSim), which is based on calculating the similarity between planted $p_*(n|k_*)$ and detected $p(n|k)$ by 
$\sum_{n=1}^N \min \left( p_*(n|k_*), p(n|k) \right)$ 
(see Methods for detailed definition of MaxSim). 
Ball-Karrer-Newman's stochastic block model (BKN’s SBM) 
%%%%% citation
\cite{Ball2011}, 
%%%%%
which is one of few existing methods that can detect pervasive communities, was chosen as baseline. For MDMC, since the number of detected communities is controlled by parameter $\alpha$ (Fig. \ref{Fig1}d), MaxSim was calculated as a function of $\alpha$ (Fig. \ref{Fig2}b). BKN's SBM has no such parameter and is required to predetermine the number of communities to which the network should be decomposed. We therefore examined BKN's SBM for $K=$10, 20 and 30. Note that $K=$10 is consistent with the number of planted communities. MaxSim given by MDMC for an extensive range of $\alpha$ surpasses that given by BKN's SBM for any $K$, indicating that MDMC well performs pervasive community detection.  

Third, we demonstrate system analyses by MDMC using real brain networks. This was first examined using a network of 10 areas in mouse visual cortex constructed from connectome data of Allen Brain Atlas (Fig. \ref{Fig2}c) 
%%%%% citation
\cite{AllenBrainAtlas, Knox2019}. 
%%%%%
Links of this network represent connection strength between areas, which are directed and weighted. MDMC for $\alpha=$0.1 detected two pervasive communities (Fig. \ref{Fig2}d, top). One (blue) and the other (red) are biased ventrally and dorsally, respectively. In both communities, VISp, the largest area centrally located, is ranked first. This makes sense because VISp is the initial gate of the visual cortex for visual sensory input, from which signals are distributed ventrally as well as dorsally. Visual cortical networks of humans and primates are known to functionally segregate to ventral and dorsal pathways, which are engaged in object recognition and contextual processing, respectively. However, whether visual cortical networks of rodents such as mice have similar functional segregation remains unclear
%%%%% citation
\cite{Wang2012}. 
%%%%%
MDMC has identified two soft-overlapping regions as pervasive communities, either of which is dorsally or ventrally biased, suggesting that the mouse visual cortical network is also functionally segregated in similar ways as for humans and primates. For $\alpha$=0.05, the network is decomposed into three pervasive communities (Fig. \ref{Fig2}d, middle): The two inherit the ventral and dorsal pathways; the third one (green) in which VISp is by far dominant is likely to gate the visual cortex. For $\alpha$=0.01, the fourth one (yellow) emerges, where areas medially located and limbic to VISp are highly ranked; this community is likely to bridge ventral and dorsal pathways (Fig. \ref{Fig2}d, bottom). 
%
%%%%%%%%%%%%%%%%%%%% Figure 2
%%%%%%%%%%%%%%%%%%%%
%

\begin{figure}[ht]
\begin{center}
%\fbox{\rule{60mm}{0mm}\rule{0mm}{40mm}}
\includegraphics[width=120mm]{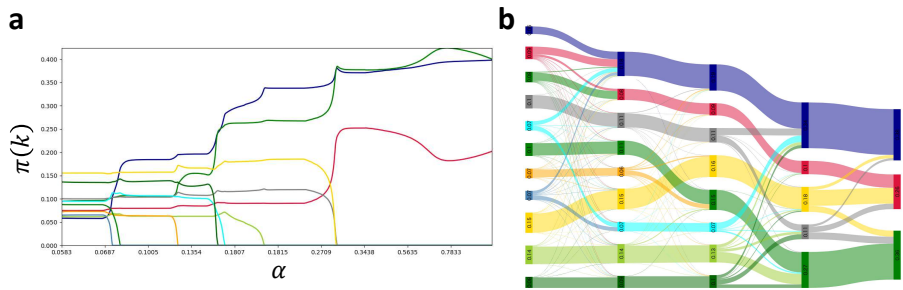}
\caption{
{\small 
Hierarchical organization of pervasive communities of the mouse whole brain network. (a) Each layer of hierarchy emerges through a series of discrete phase transitions induced by quasi-static increase in $\alpha$. (b) Sankey diagram illustrating parent-child relationships in the hierarchical organization of pervasive communities. 
The width of each band in the diagram shows the amount of flow of probability from one community at one layer to one community at the next layer.
}
}
\label{Fig3}
\end{center}
\end{figure}

The findings that the resolution of decomposition into communities changes with $\alpha$ (Fig. \ref{Fig1}e) raise an intriguing question: What is the hierarchical organization of pervasive communities? To address this, we propose the following procedure: Fix $\alpha$ to a very small value and run the EM step to decompose the network into large number of small communities; then, increase $\alpha$ quasi-statically (namely, very slowly) while continuing the EM step. Applying this procedure to the mouse whole brain network
%%%%% citation
\cite{AllenBrainAtlas, Oh2014}, 
%%%%%
we observed discrete phase transition that intermittently occurred as $\alpha$ increased (Fig. \ref{Fig3}a). At each point of phase transition, the probabilities $\pi(k)$ of some communities sharply increased whereas those of some others dropped to zero, indicating that smaller communities merged to form larger communities. Within each interval bounded by one and the next phase transitions, the probabilities $\pi(k)$ stayed almost constant, indicating a relatively stable state corresponding to a specific layer of hierarchy. Notably, the hierarchical organization of pervasive communities of the whole brain network is non-tree structured, which is well documented by Sankey diagram (Fig. \ref{Fig3}b). The non-tree structure indicates that distinct functional modules share or recruit the same functional sub-modules, implying an effective and flexible architecture of information processing in the brain. For other real-world networks, we have often observed such non-tree structured hierarchy (data not shown). These suggest that non-tree structured hierarchy of pervasive communities is a general property of real-world networks. 
%The procedure was also applied to mathematically synthesized networks planted with tree structured hierarchy of communities. We observed that the planted structure was correctly recovered (data not shown), supporting that the non-tree structured hierarchy obtained for a variety of real-world networks is not an artefact. 
%
%%%%%%%%%%%%%%%%%%%% Figure 3
%%%%%%%%%%%%%%%%%%%%
%

The modularity maximization 
%%%%% citation
\cite{Newman2006a, Newman2006b, Newman2012, Lambiotte2009, Delvenne2010, Mucha2010} 
%%%%%
and the map equation 
%%%%% citation
\cite{Rosvall2007, Rosvall2008, Shaub2012, Kheirkhahzadeh2016} 
%%%%%
are prevailing community detection methods 
%%%%% citation
\cite{Lancichinetti2009, Hric2014} 
%%%%%
that also exploit random walk. However, detecting pervasive communities is out of reach of these methods 
%%%%% citation
\cite{Fortunato2016}. 
%%%%%
Moreover, they rely on greedy search such as the Louvain method
%%%%% citation
\cite{Blondel2008}.  
%%%%%
By contrast, MDMC detects pervasive communities by using a more theoretically principled, probabilistic machine-learning approach. Computational cost of MDMC for a given resolution scales as $~O(KL)$, which means that MDMC belongs to the fastest class of algorithms to detect pervasively overlapping communities.

In this short article, we have focused on theoretical foundation of MDMC and demonstrating its capability of detecting pervasive communities using a limited spectrum of networks. More extensive evaluation using a wider spectrum of networks and comparison with other community detection methods will be discussed in forthcoming studies.

\section*{Methods}

\subsection*{EM algorithm}
We have assumed in the main text that the `global' probability $p(l)$ spreading over the entire network is decomposed as a mixture of `local' probabilities $p(l|k)\ \left(k=1,\ \cdots,\ K\right)$ as illustrated in Fig. 1a and expressed by Eq. (\ref{Eq_MDMC}). Here we show detailed derivation of the EM algorithm to solves this mixture, which is given by Eqs. (\ref{Eq_E_step}), (\ref{Eq_M_step_pi}) and (\ref{Eq_M_step_p}). Let $\mathbf{A}=\left(A_{nm}\right)$ be the adjacency matrix of the network from which we wish to detect communities; $A_{nm}$ is the weight of the link from node $m$ to node $n$. In our formulation, $A_{nm}$ is not restricted to be dichotomic (e.g. 1 (connected) or 0 (disconnected)) but can take any non-negative value. If nodes $n$ and $m$ are connected by an undirected link, we set $A_{nm}=A_{mn}$. The rate for transition from node $m$ to node $n$ is defined by $T_{nm}=A_{nm}/\sum_{n'=1}^{N}A_{n'm}$. The probability $p_t(n)$ that Mr. X is at node $n$ at time $t$ evolves obeying Eq 
. (\ref{Eq_Markov_chain})  
%%%%% citation
\cite{Page1999}.  
%%%%%
If the network is ergodic (namely, connected and irreducible), iterative calculation of Eq. (\ref{Eq_Markov_chain}) leads to a unique stationary distribution $p(n)$ satisfying
\begin{equation}
p(n) = \sum_{m=1}^{N}T_{nm}p(m)\ .
\label{Eq_steady_state}
\end{equation}
The probability $p(l)$ that Mr. X is moving along link $l$ is then given by Eq. (\ref{Eq_link_prob}). 

Imagine that $D$ ‘investigators’ are distributed over the network to search for the whereabouts of Mr. X; each investigator ought to detect which ‘street’ (namely, which link) Mr. X is moving along; search for Mr. X by individual investigators is carried out independently (namely, they neither communicate nor exchange information with each other). Suppose that investigator $d$ has observed that Mr. X is moving along the link from node $m^{(d)}$ to node $n^{(d)}$. Let $\tau^{(d)}$ denote the result of this observation. The probability of $\tau^{(d)}$ conditioned that Mr. X is staying in community $k$ is modelled by the product of categorical distributions
\begin{eqnarray}
    p(\tau^{(d)}|\left\{ p_t(n|k) \right\}_{n=1}^{N}) 
    &=&
    \prod_{n=1}^{N}\left[ p_t(n|k) \right]^{\delta_{n,\ n^{(d)}}}
    \times \prod_{m=1}^{N}\left[ p_t(m|k) \right]^{\delta_{m,\ m^{(d)}}}
    \nonumber \\
    &=&
    p_t(n^{(d)}|k)p_t(m^{(d)}|k)\ ,
    \label{Eq_observation_d_prob_cond_by_community}       
\end{eqnarray}
where $\left\{ p_t(n|k) \right\}_{n=1}^{N}$ serve as parameters, satisfying $p_t(n|k)\ge 0$ and $\sum_{n=1}^{N} p_t(n|k) =1$. These parameters are upgraded to stochastic variables by introducing the conjugate prior defined as a Dirichlet distribution in the form
\begin{equation}
    p\left(
    \left\{ p_t(n|k) \right\}_{n=1}^{N} | \left\{ p_{t-1}(n|k) \right\}_{n=1}^{N}
    \right)
    \sim
    \prod_{n=1}^{N} \left[ p_t(n|k) \right]^{
    \left( \alpha \sum_{m=1}^{N} T_{nm} p_{t-1}(m|k) +1 \right) - 1}\ ,
    \label{Eq_Dirichlet_prior}
\end{equation}
where $\alpha$ is the parameter that controls the concentration of the Dirichlet distribution. At $\alpha \rightarrow +\infty$, the Dirichlet distribution is concentrated onto the point that gives the original Markov chain $p_t(n|k) = \sum_{m=1}^{N} T_{nm} p_{t-1}(m|k)$. For finite values of $\alpha\ (>0)$, $p_t(n|k)$ fluctuates around $\sum_{m=1}^{N} T_{nm} p_{t-1}(m|k)$; the smaller the value for $\alpha$, the more apart $p_t(n|k)$ deviates from $\sum_{m=1}^{N} p_{t-1}(m|k)$. Eq. (\ref{Eq_Dirichlet_prior}) thus describes a stochastic generalization of the Markov chain. 

Results of observations by $D$ investigators are gathered to give the data 
$\mathcal{D}=\left\{ \tau^{(1)},\ \cdots,\ \tau^{(D)} \right\}$. 
Note that the elements $\tau^{(1)},\ \cdots,\ \tau^{(D)}$ are independent and identically distributed (i.i.d.). Now we introduce latent variables 
$\mathbf{z}^{(d)}=\left( z_{k}^{(d)} \right)\ \left(k=1,\ \cdots, K\right)$ 
representing in which community Mr. X was staying (this is not directly observed, whereby these variables are said ``laten'') when investigator $d$ observed him. These variables are given by a 1-of-$K$ vector (just one component is unity and the $K-1$ others are zero); for instance, if Mr. X is staying in community $k'$ when observed by investigator $d$, $z_{k}^{(d)}=\delta_{kk'}$. 
The probability of $\mathbf{z}^{(d)}$ is modelled by the categorical distribution
\begin{equation}
   p\left( \mathbf{z}^{(d)}|\left\{ \pi_t(k) \right\}_{k=1}^{K} \right)
   \sim 
   \prod_{k=1}^{K}\left[ \pi_t(k) \right]^{z_{k}^{(d)}}\ ,
   \label{Eq_prob_of_z}
\end{equation}
where $\left\{ \pi_t(k) \right\}_{k=1}^{K}$ serve as the parameters of this distribution and satisfy $\pi_t(k)\ge 0$ and $\sum_{k=1}^{K} \pi_t(k) =1$. Under the i.i.d. assumption of the data $\mathcal{D}$, the joint probability is expressed as
\begin{eqnarray}
    p\left( \mathcal{D},\ \mathbf{P}_t,\ \mathbf{Z}|\left\{ \pi_t(k) \right\}_{k=1}^{K} \right)
   &\sim& 
   \prod_{d=1}^{D}\prod_{k=1}^{K}\left[ 
   \pi(k) \prod_{n=1}^{N} \left[ p_t(n|k) \right]^{\delta_{n,\ n^{(d)}}+\delta_{n,\ m^{(d)}}}
   \right]^{z_{k}^{(d)}}
   \nonumber\\  
   &\times& 
   \prod_{k=1}^{K}\prod_{n=1}^{N} 
   \left[ p_t(n|k) \right]^{\alpha \sum_{m=1}^{N} T_{nm}p_{t-1}(m|k)}
   \label{Eq_joint_prob}
\end{eqnarray}
with the notations 
$\mathbf{P}_t=\left\{\left\{ p_t(n|1) \right\}_{n=1}^{N},\ \cdots,\ \left\{ p_t(n|K) \right\}_{n=1}^{N} \right\}$ and 
$\mathbf{Z}=\left\{ \mathbf{z}^{(d)} \right\}_{d=1}^{D}$. 

The $p(n|k)$ and $\pi(k)$ can be estimated by maximizing the joint probability w.r.t. $p_t(n|k)$ and $\pi_t(k)$. To this end, Eq. (\ref{Eq_joint_prob}) is marginalized with respect to the latent variables $\mathbf{Z}$:
\begin{eqnarray}
    p\left( \mathcal{D},\ \mathbf{P}_t|\left\{ \pi_t(k) \right\}_{k=1}^{K} \right) 
    &=&
    \sum_{\mathbf{Z}} p\left( \mathcal{D},\ \mathbf{P}_t,\ \mathbf{Z}|\left\{ \pi_t(k) \right\}_{k=1}^{K} \right)
    \nonumber \\
    &\sim&
    \prod_{d=1}^{D} \left(
    \sum_{k=1}^{K} \pi_t(k) \prod_{n=1}^{N} 
    \left[p_t(n|k)\right]^{\delta_{n,\ n^{(d)}}+\delta_{n,\ m^{(d)}}}
    \right)
    \nonumber \\    
    &\times&
    \prod_{k=1}^{K}\prod_{n=1}^{N} 
    \left[p_t(n|k)\right]^{\alpha \sum_{m=1}^{N} T_{nm}p_{t-1}(m|k)}\ .
    \label{Eq_marginalized_joint_prob}    
\end{eqnarray}
Tanking the log of this, we have
\begin{eqnarray}
    \log p\left( \mathcal{D},\ \mathbf{P}_t|\left\{ \pi_t(k) \right\}_{k=1}^{K} \right) 
    =
    \sum_{d=1}^{D} \log \left(
    \sum_{k=1}^{K} 
    \pi_t(k)\prod_{n=1}^{N} \left[p_t(n|k)\right]^{\delta_{n,\ n^{(d)}}+\delta_{n,\ m^{(d)}}}
    \right)
    \nonumber \\
    +
    \sum_{k=1}^{K}\sum_{n=1}^{N} \left(
    \alpha \sum_{m=1}^{N} T_{nm} p_{t-1}(m|k)\right)\log p_t(n|k)\ .
    \label{Eq_log_marginalized_joint_prob}
\end{eqnarray}
Introducing 
$r(k|d)\ (\ge 0)$ satisfying $\sum_{k=1}^{K} r(k|d) =1$ and then   
using Jensen’s inequality, we derive
\begin{eqnarray}
    &\log& \left(
    \sum_{k=1}^{K} 
    \pi_t(k)\prod_{n=1}^{N} \left[p_t(n|k)\right]^{\delta_{n,\ n^{(d)}}+\delta_{n,\ m^{(d)}}}
    \right)
    \nonumber \\     
    &=&
    \log \left(
    \sum_{k=1}^{K} r(k|d)
    \frac{ 
    \pi_t(k)\prod_{n=1}^{N} \left[p_t(n|k)\right]^{\delta_{n,\ n^{(d)}}+\delta_{n,\ m^{(d)}}}}{r(k|d)}
    \right)
    \nonumber \\ 
    &\ge&
    \sum_{k=1}^{K} r(k|d)
    \log \left(
    \frac{ 
    \pi_t(k)\prod_{n=1}^{N} \left[p_t(n|k)\right]^{\delta_{n,\ n^{(d)}}+\delta_{n,\ m^{(d)}}}}{r(k|d)}   
    \right)\ .
    \nonumber
\end{eqnarray}
We finally obtain the lower bound $Q$ of 
$\log p\left( \mathcal{D},\ \mathbf{P}_t|\left\{ \pi_t(k) \right\}_{k=1}^{K} \right)$:
\begin{eqnarray}
    \log p\left( \mathcal{D},\ \mathbf{P}_t|\left\{ \pi_t(k) \right\}_{k=1}^{K} \right)
    \ge Q\ ,
    \label{Eq_Jensens_inequality}
\end{eqnarray}
where
\begin{eqnarray}
    Q &=& \sum_{d=1}^{D} \sum_{k=1}^{K} r(k|d) 
    \left[
    \log \pi_t(k) + \sum_{n=1}^{N} \left(\delta_{n,\ n^{(d)}}+\delta_{n,\ m^{(d)}}\right)
    \log p_t(n|k) - \log r(k|d)
    \right]
    \nonumber \\
    &+&
    \sum_{k=1}^{K} \sum_{n=1}^{N} 
    \left(\alpha \sum_{m=1}^{N} T_{nm} p_{t-1}(m|k)\right)
    \log p_t(n|k)\ .
    \label{Eq_Q}
\end{eqnarray}
Maximization of the joint probability (\ref{Eq_joint_prob}) can therefore be substituted with maximization of this lower bound. 

Maximizing $Q$ with respect to $r(k|d)$ under the constrain $\sum_{k=1}^{k} r(k|d) =1$ gives the E-step:
\begin{eqnarray}
    r(k|d)
&=&
 \frac{\pi_t(k) \prod_{n=1}^{N} \left[ p_t(n|k) \right]^{\delta_{n,\ n^{(d)}}+\delta_{n,\ m^{(d)}}}}
    {\sum_{k=1}^{K} \pi_t(k) \prod_{n=1}^{N} \left[ p_t(n|k) \right]^{\delta_{n,\ n^{(d)}}+\delta_{n,\ m^{(d)}}}}
\nonumber \\
&=&
    \frac{\pi_t(k) p_t(n^{(d)}|k)p_t(m^{(d)}|k)}
    {\sum_{k=1}^{K} \pi_t(k) p_t(n^{(d)}|k)p_t(m^{(d)}|k)}\ .
    \label{Eq_E_step_original}
\end{eqnarray}
Maximizing $Q$ with respect to $\pi(k)$ and $p_t(n|k)$ under the constraints $\sum_{k=1}^{K} \pi_t(k) =1$ and $\sum_{n=1}^{N} p_t(n|k) =1$, respectively, gives the M-step:
\begin{equation}
    \pi_t(k) = \frac{D_k}{D}\ ,
    \label{Eq_E_step_pi_original}
\end{equation}
\begin{eqnarray}
    p_t(n|k)
 &=& \frac{\alpha}{\alpha+2D_k}\sum_{m=1}^{N}T_{nm}p_{t-1}(m|k) 
\nonumber \\
    &+&
\frac{1}{\alpha+2D_k}\sum_{d=1}^{D}r(k|d)\left(\delta_{n,\ n^{(d)}}+\delta_{n,\ m^{(d)}}\right)\ ,
    \label{Eq_E_step_p_original}
\end{eqnarray}
where $D_k=\sum_{d=1}^{D} r(k|d)$. Note that $p_t(n|k)$, which has been upgraded to stochastic variables, is solved by maximum a posteriori estimate.

Suppose that $D$ is sufficiently large. Even so, the number of observation patterns is $L$, the total number of links. Let these   observation patterns be represented by 
$\left\{ \tau_l \right\}_{l=1}^{L}$.
The frequency of observing pattern $\tau_l$ can be approximated by $Dp(l) = DT_{n_l m_l}p(m_l)$. Accordingly, we can replace the second term in the right-hand-side of Eq. (20) as
\begin{equation}
    \sum_{d=1}^D r(k|d) \left(\delta_{n,\ n^{(d)}}+\delta_{n,\ m^{(d)}}\right)
    \rightarrow
    D\sum_{l=1}^{L}p(l)r(k|l)\left(\delta_{n,\ n_l}+\delta_{n,\ m_l}\right)\ ,
\end{equation}
where $r(k|l)$ is given by Eq. (\ref{Eq_E_step}). 
Setting $\tilde{\alpha}=\alpha/2D$, we finally obtain the EM algorithm given by Eqs. (\ref{Eq_E_step}), (\ref{Eq_M_step_pi}) and (\ref{Eq_M_step_p}) in the main text, where the ornament tilde is removed for brevity.

Initial conditions of the EM step are set as follows: $\pi_0(k)\ (\ge0)$ and $p_0(n|k)\ (\ge0)$ are chosen randomly so that $\sum_{k=1}^{K}\pi(k)=1$ and $\sum_{n=1}^{N}p_0(n|k)=1$; using these $\pi_0(k)$ and $p_0(n|k)$, we define $r(k|l)$ by Eq. (\ref{Eq_E_step}). With these initial conditions, the EM step by Eqs. (\ref{Eq_E_step}), (\ref{Eq_M_step_pi}) and (\ref{Eq_M_step_p}) is iterated for a predefined number of times. We can also more elaborately define the convergence criteria to stop the EM step, but in the present study we have just heuristically defined the number of iterations. Several hundreds of iterations are normally enough to gain well convergent results, but more iterations are sometimes necessary especially when communities to be detected are non-clique like 
%%%%% citation
\cite{Shaub2012}.  
%%%%%
    
\subsection*{Teleportation: prescription for directed networks}
If the network is directed, there might be some nodes without links from them (so called ``dangling'' nodes). If so, the Markov chain has no ergodic stationary state because the probability is eventually congested upon these dangling nodes. To recover the ergodic property, we follow the prescription once proposed for the PageRank algorithm
%%%%% citation
\cite{Page1999}.  
%%%%%
We suppose that Mr. X teleports from the current node to any node with probability $\rho$; especially when Mr. X reaches a dangling node, he teleports to any node with probability unity. These processes are implemented by replacing $T_{nm}$ with
\begin{equation}
    (1-\rho)T_{nm}+(1-\rho)\frac{1}{N}I_m+\rho\frac{1}{N}\ , 
    \label{Eq_teleportation}
\end{equation}
where $I_m=1$ if node $m$ is a dangling node and $I_m=0$ otherwise. Since Mr. X is unobservable during teleportation, $\sum_{l=1}^{L}p(l)$ becomes less than unity. To resume summed-up-to-unity, we redefine 
\begin{equation}
    p(l)\leftarrow \frac{p(l)}{\sum_{l=1}^{L} p(l)}\ .
\end{equation}
The above prescription is used for the mouse visual cortical network (Fig. 2) and the mouse whole brain network (Fig. 3). 

\subsection*{Extracting hierarchical organization of pervasive communities}
MDMC has a single parameter $\alpha$. Having observed that this parameter controls the resolution of decomposing the network into communities (Fig. 1e), we sought to derive hierarchical organization of communities by making use of this property. 
Specifically, $\alpha$ is fixed to a very small value $\alpha_{\mathrm{ini}}$ and the EM step is run to decompose the network into large number of small communities; then, at time step $t_{\mathrm{ini}}$, $\alpha$ starts to quasi-statically (namely, very slowly) increase while the EM step is continued; $\alpha$ is thus increased until it reaches $\alpha_{\mathrm{fin}}$ at time step $t_{\mathrm{fin}}$. These processes gradually reduced the resolution of decomposition, whereby hierarchical structure emerged from the bottom through a series of discrete phase transitions (Fig. 3). The increment of $\alpha$ per every time step should be taken smaller for finer resolution. To implement this, we adopted the flowing schedule for changing $\alpha$:
\begin{equation}
    \alpha (t) = \alpha_{\mathrm{ini}}
    \left(
    \frac{\alpha_{\mathrm{fin}}}{\alpha_{\mathrm{ini}}}
    \right)^{(t-t_{\mathrm{ini}})/(t_{\mathrm{fin}}-t_{\mathrm{ini}})}\ .
    \label{Eq_cooling_schedule}
\end{equation}

\subsection*{Identifying parent-child relationships between pervasive communities}
Parent-child relationships in the hierarchical organization of pervasive communities are determined in the following way. Let $p^{(h)}(k|n)$ denote the belonging of node $n$ to community $k$ at layer $h$, which is specifically defined by the value for $p(k|n)$ at $\alpha = (\alpha_{h-1 \rightarrow h}+\alpha_{h \rightarrow h+1})/2$. Here, $\alpha_{h-1 \rightarrow h}$ is the value for $\alpha$ at which the discrete phase transition from layer $h-1$ to $h$ occurred. Similarly, $\pi^{(h)}(k)$ is defined to denote the probability of community $k$ at layer $h$. 
The variation of $p(k|n)$, the belonging of node $n$ to community $k$, from layer $h$ to layer $h+1$ is 
\begin{equation}
    \Delta p^{(h \rightarrow h+1)}(k|n)=p^{(h)}(k|n)-p^{(h+1)}(k|n)\ .
    \label{Eq_variation}
\end{equation}
The amounts of flow-in and flow-out of $p(k|n)$ from layer $h$ to layer $h+1$ are $\max\left(\Delta p^{(h \rightarrow h+1)}(k|n),\ 0\right)$ and $\max\left(-\Delta p^{(h \rightarrow h+1)}(k|n),\ 0\right)$, respectively. Therefore, the flow from $p^{(h)}(k'|n)$ to $p^{(h+1)}(k|n)$ is given by
\begin{eqnarray}
    f\left(p^{(h)}(k'|n) \rightarrow p^{(h+1)}(k|n)\right)
    &=&\frac{\max \left( -\Delta p^{(h\rightarrow h+1)}(k'|n) \right)}
    {\sum_{k'} \max \left( -\Delta p^{(h\rightarrow h+1)}(k'|n) \right)}
\nonumber \\
    &\times&
    \max \left( -\Delta p^{(h\rightarrow h+1)}(k|n) \right)\ .
    \label{Eq_flow}
\end{eqnarray}
Finally, the net flow from $\pi^{(h)}(k')$ to $\pi^{(h+1)}(k)$ is obtained by marginalizing Eq. (\ref{Eq_flow}) w.r.t. $n$:
\begin{equation}
    f\left(\pi^{(h)}(k') \rightarrow \pi^{(h+1)}(k)\right) 
    =
    \sum_{n=1}^{N} p(n) f\left(p^{(h)}(k'|n) \rightarrow p^{(h+1)}(k|n)\right)\ .
    \label{Eq_net_flow}
\end{equation}
Merges or splits of these flows across layers can be expressed by the Sankey diagram, which well documents parent-child relationships in hierarchical organization of pervasive communities (Fig. 3). 

\subsection*{Benchmark networks planted with pervasive communities}
Pervasive community detection was evaluated using benchmark networks planted with pervasive communities. These networks were synthesized by a specific type of stochastic block models, which has been proposed by Ball, Karrer and Newman. 

\subsubsection*{\it Ball-Karrer-Newman’s stochastic block model}
Ball-Karrer-Newman’s stochastic block model (BKN’s SBM) 
%%%%% citation
\cite{Ball2011} 
%%%%%
defines the probability of generating a network with the adjacency matrix $\mathbf{A}=\left(A_{nm}\right)$ by Poisson distribution in the form 
\begin{equation}
    p\left( \mathbf{A} \right)
    =
    \prod_{n,\ m=1}^{N} \left[
    \frac{\left(\sum_{k=1}^K \theta_{nk}\theta_{mk} \right)^{A_{nm}}}{A_{nm}!}
    \exp\left(
    -\sum_{k=1}^K \theta_{nk}\theta_{mk}
    \right)
    \right]\ .
    \label{Eq_BKN_SBM}
\end{equation}
Here, $\theta_{nk}$ is a parameter representing the “propensity” of node $n$ to block $k$ and taking a continuous non-negative value, whereby delineating the pervasive structure of block $k$; $\sum_{k=1}^K \theta_{nk}\theta_{mk}$ is the rate for a Poisson event of generating a link between nodes $n$ and $m$. BKN’s SBM thus generates networks planted with pervasively structured blocks (namely, pervasive communities). Since $\sum_{k=1}^K \theta_{nk}\theta_{mk}$ is symmetric between $n$ and $m$, BKN’s SBM is basically applicable for undirected networks. In the rest of this section, therefore, we assume that the adjacency matrix is symmetric ($A_{nm}=A_{mn}$).

BKN’s SBM can also be used for pervasive community detection. 
This is achieved by inferring $\theta_{nk}$ for the adjacency matrix $\mathbf{A}=\left(A_{nm}\right)$ of a given network. Ball-Karrel-Newman derived the EM algorithm to solve $\theta_{nk}$, as follows: 

\noindent
{\it E-step}
\begin{equation}
    q_{nm}(k) = \frac{\theta_{nk}\theta_{mk}}{\sum_{k=1}^K \theta_{nk}\theta_{mk}}\ ,
    \label{Eq_BKN_SBM_E_step}
\end{equation}
\noindent
{\it M-step} 
\begin{equation}
    \theta_{nk}=\frac{\sum_{m=1}^N A_{nm}q_{nm}(k)}{\sqrt{\sum_{m=1}^N A_{nm}q_{nm}(k)}}\ .
    \label{Eq_BKN_SBM_M_step}    
\end{equation}
Thus, we can solve $\theta_{nk}$ by iteratively calculating Eqs. (\ref{Eq_BKN_SBM_E_step}) and (\ref{Eq_BKN_SBM_M_step}). 

\subsubsection*{\it MDMC for a specific case ($\alpha$=0) is equivalent to BKN’s SBM}
Here we demonstrate that BKN’s SBM is a specific instance of MDMC. Setting $\alpha=0$ in Eq. (\ref{Eq_M_step_p}), we have
\begin{equation}
    p_t(n|k) = \frac{1}{2\pi_t(k)}
    \sum_{l=1}^L p(l) r(k|l) \left(\delta_{n,\ n_l}+\delta_{n,\ m_l}\right)\ .
    \label{Eq_alpha_0_M_step_p}
\end{equation}
For undirected networks ($A_{nm}=A_{mn}$), the steady state distribution $p(n)$ can be given in the analytic form 
%%%%% citation
\cite{Lambiotte2009, Delvenne2010, Mucha2010}:  
%%%%%
\begin{equation}
    p(n)=\frac{\sum_{m=1}^N A_{nm}}{2L}=\frac{\sum_{m=1}^N A_{mn}}{2L}\ , 
    \label{Eq_steady_state_distribution}
\end{equation}
where $2L=\sum_{nm}^{N}A_{nm} $. This leads
\begin{equation}
    p(l)=T_{n_l m_l}p(m_l) = \frac{A_{n_l m_l}}{\sum_n A_{n m_l}}
    \frac{\sum_n A_{n m_l}}{2L}=\frac{A_{n_l m_l}}{2L}=\frac{A_{m_l n_l}}{2L}\ .
    \label{Eq_steady_state_distribution_link}
\end{equation}
Eq. (\ref{Eq_alpha_0_M_step_p}) can therefore be rewritten as
\begin{equation}
    p_t(n|k) = \frac{1}{2L\pi_t(k)} \sum_{m=1}^N A_{n_l m}r(k|l)\ .
    \label{Eq_alpha_0_M_step_p2}
\end{equation}
Now define $q_{n_l m_l}(k)$ and $\theta_{nk}$ in terms of $r(k|l)$, $p_t(n|k)$ and $\pi_t(k)$ as
\begin{equation}
    q_{n_l m_l}(k) = r(k|l)\ ,
    \label{Eq_q}
\end{equation}	
\begin{equation}
    \theta_{nk}=\sqrt{2L\pi_t(k)}p_t(n|k)\ .
    \label{Eq_theta}
\end{equation}
Rewriting Eqs. (\ref{Eq_alpha_0_M_step_p2}) and (\ref{Eq_E_step}) in terms $q_{nm}(k)$ of Eq. (\ref{Eq_q}) and $\theta_{nk}$ of Eq. (\ref{Eq_theta}) just gives the EM algorithm of BKN's SBM given by Eqs (\ref{Eq_BKN_SBM_E_step}) and (\ref{Eq_BKN_SBM_M_step}). 

MDMC defines the probability of observations $\mathcal{D}=\left\{ \tau^{(1)},\ \cdots,\ \tau^{(D)}\right\}$ as
\begin{equation}
   p\left(\mathcal{D}\right)=
   \prod_{d=1}^D \left[\sum_{k=1}^K \pi(k) p(n^{(d)}|k)p(m^{(d)}|k) \right]\ .
   \label{Eq_data_prob}
\end{equation}
Suppose that the number of times at which Mr. X is observed moving along link $l$ is $A_{n_l m_l}$. The result of $\sum_{n,\ m=1}^N A_{nm}=2L$ times of observation can therefore be expressed as $\left\{A_{nm}\right\}_{n,\ m=1}^N$. Note here that $A_{nm}=0$ if nodes $n$ and $m$ are unconnected. From Eq. (\ref{Eq_data_prob}), we can derive the probability of $\left\{A_{nm}\right\}_{n,\ m=1}^N$ given by a multinomial distribution in the form
\begin{equation}
    p\left(\mathbf{A}
    |\sum_{n,\ m=1}^N A_{nm}=2L
    \right)
    =
    \frac{(2L)!}{\prod_{n,\ m=1}^N A_{nm}!}
    \prod_{n,\ m=1}^N \left[
    \sum_{k=1}^K \pi(k) p(n|k)p(m|k)
    \right]^{A_{nm}}\ .
\end{equation}
This can be further arranged as
\begin{eqnarray}
    p\left(\mathbf{A}
    |\sum_{n,\ m=1}^N A_{nm}=2L
    \right)
    &=&
 \prod_{n,\ m=1}^N
    \frac{
   \frac{\left[
    2L\sum_{k=1}^K \pi(k) p(n|k)p(m|k)
    \right]^{A_{nm}}}{A_{nm}!}
    }
    {\frac{(2L)^{2L} \exp(-2L)}{(2L)!}}
\nonumber \\
&\times& 
    \exp\left(
    -2L\sum_{k=1}^K \pi(k) p(n|k)p(m|k)
    \right)\ .
\end{eqnarray}
Noticing 
$p(\sum_{n,\ m=1}^N A_{nm}=2L)=\frac{(2L)^{2L} \exp(-2L)}{(2L)!}$, 
we finally have
\begin{eqnarray}
    p\left( \mathbf{A}\right)
    &=&
    p\left( \mathbf{A}|\sum_{n,\ m=1}^N A_{nm}=2L\right)
    p(\sum_{n,\ m=1}^N A_{nm}=2L)
    \nonumber \\
    &=&
    \prod_{n,\ m=1}^N\frac{\left[
    2L\sum_{k=1}^K \pi(k) p(n|k)p(m|k)
    \right]^{A_{nm}}}{A_{nm}!}
\nonumber \\
&\times&
    \exp\left(
    -2L\sum_{k=1}^K \pi(k) p(n|k)p(m|k)
    \right)\ .
    \label{Eq_generating_benchmark_NWs_prob}
\end{eqnarray}
Expressed in terms of $q_{nm}(k)$ and $\theta_{nk}$, this turns to become Eq. (\ref{Eq_BKN_SBM}). Thus, we conclude that MDMC for $\alpha=0$ and BKN’s SBM are equivalent. 

\subsubsection*{\it Benchmark networks planted with pervasive communities}
Benchmark networks used in the present study were synthesized using BKN’s SBM, as follows. First, $\left\{p_*(n|k_*)\right\}_{n=1}^N$ and $\left\{\pi_*(k_*)\right\}$ ($k_*=1,\ \cdots,\ K_*$)
were stochastically generated so that they follow power-law distributions 
$p\left(p_*(n|k_*)\right)\sim \left[p_*(n|k_*)\right]^{-\gamma}$ 
and 
$p\left(\pi_*(k_*)\right)\sim \left[\pi_*(k_*)\right]^{-\beta}$, 
respectively. 
Here, the subscript asterisk is used to discriminate planted ones from $p(n|k_*)$ and $\pi(k_*)$ to be inferred.
Assuming the power-law distributions for 
$\left\{p_*(n|k_*)\right\}_{n=1}^N$ and $\left\{\pi_*(k_*)\right\}$ 
stems from the empirical fact that distributions of community sizes and degrees of nodes follow power laws in many of real-world networks
%%%%% citation
\cite{LFR_bench2008, LF_bench2009}. 
%%%%%
The stochastic generation of 
$\left\{p_*(n|k_*)\right\}_{n=1}^N$ and $\left\{\pi_*(k_*)\right\}$ 
was also devised so that the ratios $\max_n p_*(n|k_*)/\min_n p_*(n|k_*)$ and $\max_{k_*} \pi_*(k_*)/\min_{k_*} \pi_*(k_*)$ fall within moderate ranges.
Benchmark networks were then generated according to the probability (\ref{Eq_generating_benchmark_NWs_prob}). The parameter values for the synthesis were set as follows: $N=1000$; $K_*=10$; $\gamma=3$; $\beta=2$. 
 
\subsubsection*{\it Measuring the performance of pervasive community detection}
To quantitatively measure the performance of pervasive community detection, we have introduced the “maximum similarity (MaxSim)”. Let $K_*$ and $K$ be the number of communities planted in the benchmark network and the number of detected communities, respectively. 
%; $C_{k_*}$ and $C_k'$ denote the $k$-th planted community and the $k'$-th detected community, respectively
The similarity between planted $p_*(n|k_*)$ and detected $p(n|k)$, defined by 
$\mathrm{Sim}(k_*,\ k) =\sum_{n=1}^N \min \left(p_*(n|k_*),\ p(n|k)\right)$, 
is then calculated for all combinations of $k_*\ (k_*=1,\ \cdots,\ K_*)$ and $k\ (k=1,\ \cdots,\ K)$. MaxSim is hence given by
\begin{equation}
    \mathrm{MaxSim}=
    \sum_{k=1}^{K_*} \pi_*(k_*) 
    \left(1-\frac{\left|\pi(k_*)-\pi(\arg\max_k \mathrm{Sim}(k_*, k))\right|}{\pi(k_*)+\pi(\arg\max_k \mathrm{Sim}(k_*, k))}\right)
    \max_{k} \mathrm{Sim}(k_*, k) .
    \label{Eq_MaxRC4}
\end{equation}
MaxSim ranges from zero to unity. 
When MaxSim=1, recovery of planted pervasive communities is perfect. 
For each value of $\alpha$, MDMC is examined for 24 benchmark networks (synthesized using the same parameter values but with different random seeds) and MaxSim is averaged over these networks. Pervasive community detection by BKN’s SBM is taken as a baseline for comparison.
This competition seems advantageous to BKN's SBM because the task is to detect (namely, decode) the community structure that is generated (namely, encoded) by BKN's SBM itself. We show in the main text that, despite this, MDMC outperforms BKN's SBM.   

\subsubsection*{\it Brain network data}
We believe that brain networks are the best examples of real-world networks that have pervasive communities. Therefore, we examined two brain networks, both constructed from connectome data downloaded from Allen Mouse Brain Connectivity Atlas
%%%%% citation
\cite{AllenBrainAtlas, Knox2019, Oh2014}. 
%%%%%
One is a network of 10 areas in the mouse visual cortex: primary visual area (VISp); lateral visual area (VISl); laterointermediate area (VISli); posterolateral visual area (VISpl); postrhinal area (VISpor); posteromedial visual area (VISpm); Anteromedial visual area (VISam); Anterior area (VISa); rostrolateral visual area (VISrl); Anterolateral visual area (VISal). The other is a whole brain network of 213 cortical areas in the right hemisphere. Links of both networks are weighted and directed.

%%%%%%%%%%%%%%%%%%%%%%%%%%%%%%
% References
%%%%%%%%%%%%%%%%%%%%%%%%%%%%%%
\bibliographystyle{unsrt}  


\end{document}